\documentstyle[12pt]{article}
\begin{document}
\setlength{\baselineskip}{.7cm}
\renewcommand{\thefootnote}{\fnsymbol{footnote}}
\sloppy

\newcommand \be  {\begin{equation}}
\newcommand \bea {\begin{eqnarray} \nonumber }
\newcommand \ee  {\end{equation}}
\newcommand \eea {\end{eqnarray}}

\begin{center}
\centering{\bf Causal cascade in the stock market 
from the ``infrared'' to the ``ultraviolet''.}
\end{center}

\begin{center}
\centering{Alain Arneodo$^1$,  Jean-Fran\c{c}ois Muzy$^1$ and Didier
Sornette$^{2,3}$\\
{\it
$^1$ Centre de Recherche Paul Pascal, Av. Schweitzer, 33600 Pessac,
France\\
$^2$ Department of Earth and Space Science\\ and Institute of Geophysics
and Planetary Physics\\ University of California, Los Angeles,
California 90095\\
$^3$ Laboratoire de
Physique de la Mati\`ere Condens\'ee, CNRS URA190\\ Universit\'e des
Sciences, B.P. 70, Parc Valrose, 06108 Nice Cedex 2, France
}
}
\end{center}

\date{\today}

\vskip 1cm

{\bf Modelling accurately financial price variations is an essential step
underlying portfolio allocation optimization, derivative pricing and
hedging, fund
management and trading.
The observed complex price fluctuations guide and 
constraint our theoretical understanding of 
agent interactions and of the organization of the market. The gaussian
paradigm of
independent normally distributed price increments
\cite{Bachelier,Samuelson} has long
been known to be incorrect with many attempts to improve it. Econometric
nonlinear
autoregressive models with conditional heteroskedasticity\cite{Engle} 
(ARCH) and
their generalizations \cite{Bollerslev} 
capture only imperfectly the volatility correlations
and the
fat tails of the probability distribution function (pdf) of price
variations. 
Moreover, as far as changes in time scales are concerned, 
the so-called ``aggregation'' properties of these models 
are not easy to control.
More recently, the leptokurticity of the full pdf 
was described by a truncated ``additive''
L\'evy flight model\cite{Stanley,Arneodo} (TLF).
Alternatively, Ghashghaie {\em et al.}\cite{Peinke} proposed an 
analogy between price dynamics and hydrodynamic turbulence.

In this letter, we use wavelets to decompose the volatility of 
intraday (S\&P500) return data across scales.
We show that when investigating
two-points correlation functions of the
volatility logarithms across different time scales, one reveals
the existence of 
a causal information 
cascade from large scales (i.e. small frequencies, 
hence to vocable ``infrared'') to 
fine scales (``ultraviolet''). We quantify and visualize the
information flux across scales.
We provide a possible interpretation of our findings 
in terms of market dynamics.}

The controversial \cite{Arneodo,comment} analogy developed by
Ghashghaie {\em et al.}\cite{Peinke} 
implicitly assumes that 
price fluctuations can be described 
by a {\em multiplicative cascade}
along which, the return at a given scale $a < T$,
is given by:
\begin{equation}
   r_a(t) \equiv \ln P(t+a) - \ln P(t) = \sigma_{a}(t) u(t) \; ,
\label{proc}
\end{equation}
where $u(t)$ is some scale independent random variable, $T$
is some coarse ``integral'' time scale 
and $\sigma_a(t)$
is a positive quantity that can be
multiplicatively decomposed,
for each decreasing sequence of scales $\{a_i\}_{i=0,..,n}$ with
$a_0 = T$ and $a_n = a$, as\cite{Castaing,manu}
\begin{equation} 
 \sigma_{a} = \prod_{i=0}^{n-1} W_{a_{i+1},a_{i}} \sigma_T \; .
\label{multi}
\end{equation}
In turbulence, the field $\sigma$ is related to the
energy while in finance $\sigma$ is called the 
volatility. Recall that the volatility has fundamental importance
in finance since it provides a measure of the amplitude of price
fluctuations, hence of the market
risk. Using $\omega_a (t) \equiv \ln \sigma_a(t)$ 
as a natural variable, 
if one supposes that $W_{a_{i+1},a_i}$  
depends only on the scale ratio $a_i/a_{i+1}$, 
one can easily show,
by choosing
the $a_i$ as a geometric series $T s^n \; (s<1)$,
that eq.~(\ref{multi}) implies that the pdf
of $\omega$ at scale $a$ can 
be written as\cite{Castaing,manu} 
\begin{equation}
    p_a (\omega) = (G_{s}^{\otimes n} 
                          \otimes p_T) (\omega) \; ,
\label{pdf}
\end{equation}
where $\otimes$ means the convolution product,
$G_{s}$ is the pdf of $\ln W_{s a,a}$ and
$p_T$ is the pdf of $\omega_{T}$.
The above equation is the exact reformulation
(in log variables) of the paradigm that Ghashghaie {\em et al.}
\cite{Peinke}
used to fit foreign exchange (FX) rate data at different scales. 
In this formalism, $G$ can be proven to be the
pdf of an infinitely divisible random variable \cite{manu}
(hence $\sigma$ is called ``log-infinitely divisible'').
In ref. \cite{Peinke}, $G$ is assumed to be Normal
(the cascade is called ``log-normal'') of variance 
$- \lambda^2 \ln s $.

First, let us comment on the criticisms
raised by Mantegna and Stanley \cite{comment}.
Note that eq.~(\ref{pdf})
does not determine the shape of the pdf of the returns $r_a(t)$
{\em at a given scale} but specifies how this
pdf changes {\em across scales}. For a fixed scale,
the precise form for the pdf depends on both
$p_T$ and on the law of the variable $u(t)$
(which determines notably the sign of $r_a(t)$).
Therefore, nothing 
prevents the pdf of $r_a(t)$ to having fat tails at small scales
as observed in financial time series \cite{Peinke}.
A cascade model actually accounts for the distribution of
the volatility of returns across scales and not
for the precise fluctuations of $r_a(t)$.
The behavior of the autocorrelation
function $\overline{r_a(t)r_a(t+\tau)}$ ($\tau > a$) indeed depends on 
both the cascade variables and $u(t)$.
For example, if $u(t)$ is a white noise, there 
will be no correlation between the returns
while their absolute values (or the associated volatilies)
are strongly correlated (see below). 
This is why the shape of
the power spectrum of financial time series cannot
be invoked as an argument against a cascade model.
Moreover, as far as scaling properties of price fluctuations 
are concerned, it is easy to deduce from eq. (\ref{pdf}) that, if 
$H \ln s $ is the mean of $G_s$ and $-\lambda^2 \ln s$ its 
variance, then
the the maximum of the pdf of $\sigma_a(t)$
varies as $a^{H-\lambda^2/2}$ ($H$ plays the same role as the L\'evy
index in TLF models with $H=1/\mu$) 
while its standard deviation behaves as 
$a^{H-\lambda^2}$; these features are observed in 
both turbulence \cite{Castaing} 
($H\simeq 0.33$ and $\lambda^2 \simeq 0.03$) and finance
\cite{Peinke}  ($H \simeq 0.6$ and $\lambda^2 \simeq 0.015$).
Therefore, as advocated in ref. \cite{Peinke},
eq.~(\ref{pdf}) accounts reasonably well for 
one-point statistical properties
of financial times series. However, 
because of the relatively small statistics
available in finance, it is very difficult
to demonstrate that eq. (\ref{pdf}) is more pertinent
to fit the data than a ``truncated L\'evy'' 
distribution \cite{Stanley,Arneodo,comment}.

At this point, let us emphasize that
eq.~(\ref{multi}) imposes much more constraints on the statistics 
(it is indeed a model !)
than eq.~(\ref{pdf}) that only refers to one point statistics.
The main difference between 
the {\em multiplicative} cascade model and the 
truncated L\'evy {\em additive} model is that
the former predicts strong
correlations in the volatility while the latter
assumes no correlation.
It is then tempting to compute the correlations of the log-volatility
$\omega_a$ at different time scales $a$.
For that purpose, we use a natural tool to perform time-scale
analysis, the {\em wavelet transform} (WT).
Wavelet analysis has been introduced
as a way to decompose signals in
both time and scales \cite{wavelets}. 
The WT of $f(t) = \ln P(t)$ is defined as:

\begin{equation}
T_{\psi}[f](t,a) \equiv \frac{1}{a} \int_{-\infty}^{+\infty}f(y)\psi \left(\frac{y-t}{a}\right)dy,
\end{equation}
where $t$ is the time parameter, $a$ ($>$0) the scale parameter 
and $\psi$ the analyzing wavelet.
Note that for $\psi(t) = \delta(t-1) - \delta(t)$, 
$T_{\psi}[f](t,a)$ is nothing but the 
return $r_a(t)$. However, in general, $\psi$ is 
choosen to be well localized in both time and frequency,
so that the scale $a$ can be interpreted as an 
inverse frequency.
Moreover, if $\psi$ has at least two vanishing moments and 
$\chi$ is a bump function with $||\chi||_1 = 1$,
then, the {\em local} volatility at scale $a$ and
time $t$ can be defined as \cite{Tcham}
$\sigma_a^2(t) \equiv a^{-3} \int \chi((b-t)/a) |T_{\psi}(b,a)|^2 db$.
Actually, thanks to the time-scale properties of the 
wavelet decomposition \cite{wavelets}, when summing $\sigma_a^2(t)$
over  time and scale, one recovers the total square derivative
of $f$: $\Sigma = \int\int \sigma_a^2(t) dt da = \int |df/dt|^2 dt$.

In Fig.~1 are shown 3 time series for which
we study the increment time correlations.
Fig.~1(a) represents the logarithm of 
the $S\&P500$ index.
The corresponding ``volatility walk'',
$v_a(t) = \sum_{i=0}^t \omega_a(i)$ 
is represented in Fig.~1(b).
Fig.~1(c) is the same as Fig.~1(b) but after having randomly 
shuffled 
the increments $\ln P(i+1) - \ln P(i)$
of the signal in Fig.~1(a).
Fig.~1(b) clearly demonstrates 
the existence of important long-range positive 
temporal correlations  
in the volatilities of $S\&P500$ returns. 
Moreover, the statistics of $\omega_a(t)$ are found to be
nearly gaussian. 
However, the volatility walk for the ``shuffled S\&P500''
looks very much like a Brownian motion with uncorrelated increments. 
This observation is sufficient
to discard any additive (like TLF) model which intrinsically 
fails to account for the strong correlations
observed in $\omega_a(t)$.
The correlation function
$C^r_1(\Delta t) = \overline{r_1(t)r_1(t+\Delta t)} - \overline {r_1(t)}^2$
shown in Fig.~1(a'),
confirms the well-known fact that
there are no correlations between the returns 
(except at a very small time lag as illustrated in the inset). 
However, the difference is striking in Fig. 1(b') where
the correlation function of the volatility walk
$C^{\omega}_a(\Delta t) = \overline {\omega_a(t)\omega_a(t+\Delta t)} - 
\overline {\omega_a(t)}^2$
remains as large as 5\% up to time
lags corresponding to about two months.
In contrast, the correlation function associated to 
the shuffled time series in Fig.~1(c') is within the noise level.

From the modelling of fully developed turbulent flows and 
fragmentation processes, random multiplicative
cascade models are well known to generate 
long-range correlations \cite{K62,novikovstewart,Mandelbrot}.
We now explore whether this concept could be useful for understanding
the observed 
long-range correlations of the volatility 
(and not of the price increments, which
makes turbulence and financial markets drastically different). 
To fix ideas, let us consider a specific realization of
a process satisfying eq.~(\ref{multi}).
Consider the largest time scale $T$ of
the problem. We then assume that the volatility at time scale $T$ influences
the volatility of the two subperiods 
of length ${T \over 2}$ by random factors
equal respectively to
$W_0$ and $W_1$. In turn, each volatility over ${T \over 2}$ influences
the two
subperiods of length ${T \over 4}$ by random 
factors $W_{00}$ and $W_{10}$
for the
first sub-period and $W_{01}$ and $W_{11}$ for the second one. 
The cascade
process is assumed to continue 
along the time scales until the shortest tick time scale 
(see ref. \cite{manu} for rigourous definitions and properties).
The simplest assumption is that the factors $W$ are i.i.d.
variables with log-normal distribution of mean $-H\ln 2$ 
and variance $\lambda^2 \ln 2$.
It is then easy to show that the correlation function 
averaged over a period of length $T$,
$C^{\omega}_a(\Delta t) = T^{-1} \int_0^T
\left(  \left< \omega_a(t)\omega_a(t+\Delta t) \right> - 
 \left< \omega_a(t)\right>^2 \right) dt$, 
can be written as 
\begin{equation}
  C^{\omega}_a(\Delta t) = \lambda^2 (1 - \log_2 {\Delta t\over T} - 2 
{\Delta t\over T}) ~,  
\label{corr}
\end{equation}
for $a \leq \Delta t \leq T$ ($\left< . \right>$ means mathematical
expectation). Here, our goal is to show that the basic
ingredients of this simple cascade model are sufficient 
to rationalize most of the features observed on the volatility
correlations at different scales
(note that one could improve this description by
taking into account mutual influences
of volatilities at a given scale and
the possible ``inverse cascade'' influence of fine scales
on larger ones). 
For $\lambda^2 \simeq 0.015$ obtained independently from the
fit of the pdf's \cite{Peinke}, 
eq.~(\ref{corr}) provides 
a very good fit of the data (Fig 1(b')) 
for the slow decay of the correlation function with only
one adjustable parameter $T \simeq 3 $ months.
Let us note that 
$C^{\omega}_a(\Delta t)$ can be equally well fitted by a power law $\Delta t^{-\alpha}$
with
$\alpha \approx 0.2$. In view of the small value of $\alpha$, this is
undistinguishable from a logarithmic decay.
Moreover, eq. (\ref{corr}) predicts that 
the correlation function $C^{\omega}_a(\Delta t)$ should 
not depend of the scale $a$ provided $\Delta t > a$. 
In Fig. 2, 
$C^{\omega}_a(\Delta t)$ are plotted versus $\ln(\Delta t)$ for 
various scales $a$ corresponding to 30, 120 and 480 min.
As expected, all the data collapse on a single curve
which is nearly linear up to 
some integral time of the order of 3 months.

Let us point out that 
volatility at large time intervals that cascades to smaller scales cannot
do so instantaneously. From causality properties of 
financial signals, the ``infrared'' towards
``ultraviolet'' cascade must manifest itself in a
time asymmetry of the cross-correlation coefficients
$C^{\omega}_{a_1,a_2}(\Delta t) \equiv  \mbox{var}(\omega_{a_1})^{-1}
\mbox{var}(\omega_{a_2})^{-1}(
\overline{\omega_{a_1}(t)\omega_{a_2}(t+\Delta t)} -
\overline{\omega_{a_1}(t)} \; \overline{\omega_{a_2}(t)})$; in particular, 
one expects that $C^{\omega}_{a_1,a_2}(\Delta t) > C^{\omega}_{a_1,a_2}(-\Delta t)$
if $a_1 > a_2$ and $\Delta t > 0$.
From the near-Gaussian properties of $\omega_a(t)$, the mean mutual
information of the variables $\omega_a(t+\Delta t)$ and
$\omega_{a+\Delta a}(t)$ reads\,: 
\begin{equation}
I_a(\Delta t,\Delta a) = -0.5 \log_2 \left( 1-
(C^{\omega}_{a,a+\Delta a}(\Delta t))^2 \right) \; .  \label{info}
\end{equation}
Since the process is causal, this quantity can be interpreted
as the information contained in $\omega_{a+\Delta a}(t)$
that propagates to $\omega_a(t+\Delta t)$.
In Fig.~3, we have computed  
$I_{a}(\Delta t,\Delta a)$ for the S\&P500 index (top) and
its randomly shuffled version (bottom). 
One can see on the bottom picture that there is 
no well defined structure that emerges from the noisy background. 
Except in a small domain at small scales around $\Delta t = 0$, 
the mutual information is in the noise level as expected for
uncorrelated variables.  
In contrast, two features are clearly visible on the top
representation.
First, the mutual information at different scales is mostly
important for equal times. This is not so surprising since
there are strong localized structures in the signal that are
``coherent'' over a wide range of scales.
The extraordinary new fact is the appearance of 
a non symmetric propagation cone of information showing that the
volatility a large scales influences causally (in the future) 
the volatility at shorter scales. Although one can also detect   
some information that propagates from  
past fine to future coarse scales, 
it is clear that this 
phenomenon is weaker than past coarse/future fine flux 
(the fact that the former one exists anyway suggests 
that a more realistic cascading process should 
include the causal influence of short time scales on larger ones).
Figure 3 is thus a clear demonstration of the pertinence of 
the notion of a cascade in market dynamics.
Similar features have been found on Foreign Exchange rates.

There are several mechanisms that can be invoked to rationalize our
observations,
such as the heterogeneity of traders and their different 
time horizon \cite{Muller}
leading to an ``information'' 
cascade from large time scales to short time
scales,  
the lag between stock market fluctuations and long-run movements in
dividends \cite{Barsky}, the effect of the regular release (monthly,
quarterly) of major economic indicators 
which cascades to fine time scale.
Correlations of the volatility have been known for a 
while and have been partially modelled by mixtures of distributions
\cite{mixture}, ARCH/GARCH models \cite{Engle} and their 
extensions \cite{Bollerslev}.
However, as pointed out in the introduction, because
they are constructed to fit the fluctuations at a given
time interval, these models
are not adapted to account for the above described 
multi-scale properties of financial time series.
We have performed the same correlation analysis for simulated GARCH(1,1)
processes and obtained structureless pictures similar to 
the one corresponding to the shuffled S\&P500 in Fig.~3(b). 
More recently, Muller {\em et al.} \cite{Muller} have proposed the
HARCH model in which the variance 
at time $t$ is a function of the realized variances
at different scales. 
By construction, this model captures the lagged
correlation of the volatility from the large to the small time scales. 
However, it does not contain the notion of cascade
and involves only a few time scales. Moreover, it suffers
from the same defficiencies as ARCH-type models concerning
the difficulties to control and interpret parameters 
at different scales.

Putting together the
evidence provided by the logarithmic decay of the volatility correlations
and the
volatility cascade from the infrared to the ultraviolet, 
we have revisited the analogy with turbulence, albeit on the {\it
volatility} and not on the price variations. 
Another very promising prospect consists in building 
ARCH-type processes on orthogonal wavelets basis. 
This work is in current progress.
The present understanding with such models will allow us to
calculate improved risk prices 
such as options, for instance using the functional
formalism of ref. \cite{BS} 
well-adapted to deal with pdf's of the form (\ref{pdf}). 

 \noindent
{\bf Acknowledgments.} We acknowledge useful discussions with E. Bacry and 
U. Frisch.

\newpage

\noindent
{\bf \large Figure Captions} \\

\noindent
{\bf Figure 1:} (a) Time evolution of $\ln P(t)$, where $P(t)$ is the 
S\&P500 index, sampled with a time resolution $\delta t = 5$ min in 
the period October 1991-February 1995. The data have been preprocessed 
in order to remove ``parasitic'' daily oscillatory 
effects. (b) The corresponding
``volatility walk'', $v_a(t) = \sum_{i=0}^t \omega_a(i)$, as computed 
with a compactly supported spline wavelet\cite{wavelets} for
$a = 4$ ($\simeq 20$ min). (c) $v_a(t)$ computed after having randomly 
shuffled the increments of the signal in (a). (a') The 5 min return correlation
function $C_1^r(\Delta t)$ versus $\Delta t$ from $0$ to $20$ min. (b')
The correlation function  $C_a^{\omega}(\Delta t)$ of the
log-volatility of the S\&P500 at scale $a = 4$ ($\simeq$ 20 min); the
solid line corresponds to a fit of the  data using eq.~(\ref{corr})
with $\lambda^2 = 0.015$ and $T \simeq 3$ months. (c') same as in (b')
but for the randomly shuffled S\&P500 signal. In (a'-c') the dashed
lines delimit the 95\% confidence interval.\\

\noindent
{\bf Figure 2:} The correlation function $C_a^{\omega}(\Delta t)$ of 
the log-volatility of the S\&P500 index is plotted versus $\ln \Delta t$
for various scales $a$ corresponding to 30 ($\circ$), 120 ($\times$)
and 480 ($\triangle$) minutes. All the data collapse on a same curve 
which is almost linear up to an integral time scale 
$T \simeq$ 3 months ($\ln T = 8.6$). According to eq.~(\ref{corr}),
from the slope of this straight line, one gets an estimate of 
the parameter $\lambda^2 \simeq 0.015$.\\

\noindent
{\bf Figure 3:} The mutual information $I_a(\Delta t, \Delta a)$ (eq.~(\ref{info}))
of the variables $\omega_{a}(t+\Delta t)$ and $\omega_{a+\Delta a}(t)$ is 
represented in the $(\Delta t, \Delta a)$ half-plane (5 min units); 
the time lag
$\Delta t$ spans the interval $[-2048, 2048]$ while 
the scale lag $\Delta a$ ranges from $\Delta a = 0$ (top) to 1024 (bottom).
The amplitude of $I_a(\Delta t, \Delta a)$ is coded from black for zero values 
to red for maximum positive values (``heat'' code), independently at
each scale lag $\Delta a$. (a) S\&P500 index; (b) its randomly shuffled
increment version. Note that, for middle scale lag values, the maxima
(red spots) of the mutual information in (a) are 2 order of magnitude
larger than the  corresponding maxima in (b).      
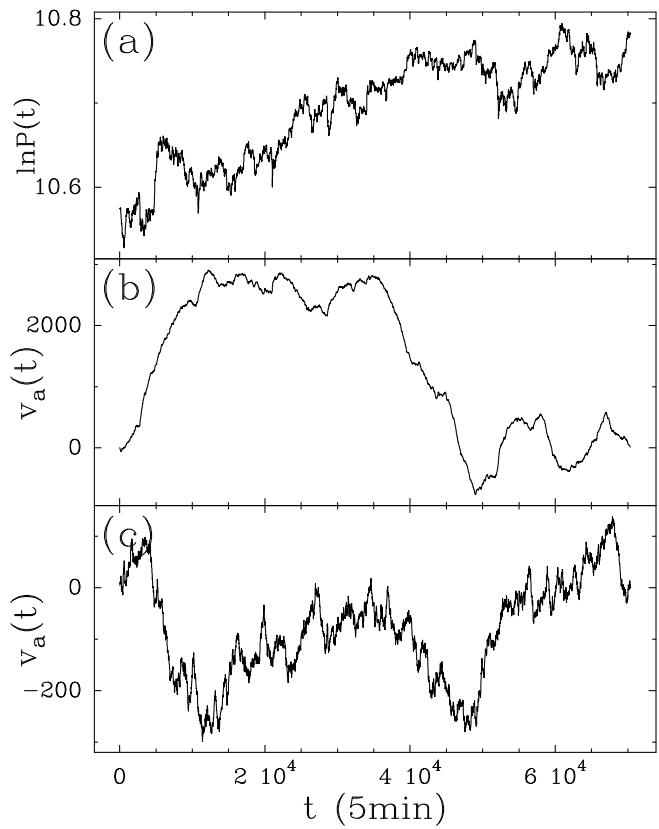
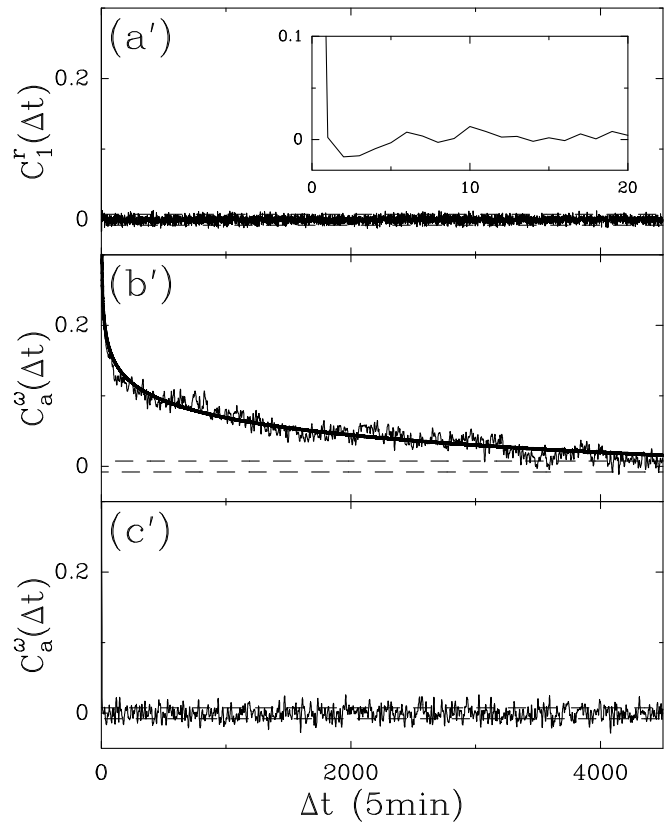
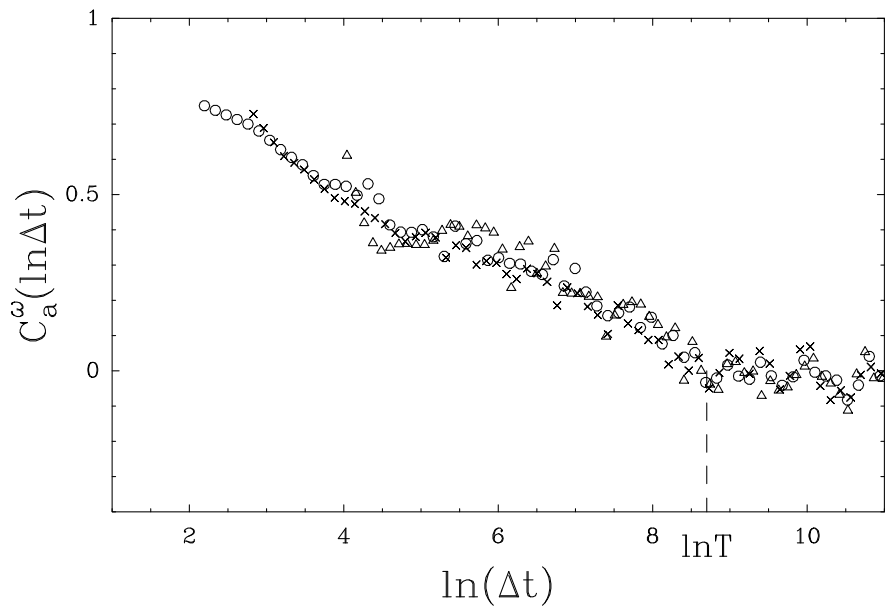
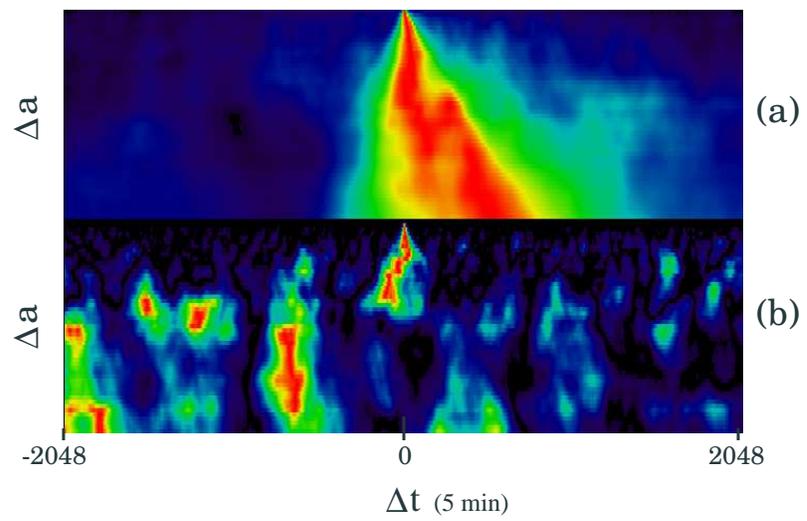
\end{document}